\documentclass[a4paper,11pt]{article}
\usepackage{pos}

\title{Phase diagram of QCD in strong background magnetic field}

\author[a]{Massimo D'Elia}
\author*[a]{Lorenzo Maio}
\author[b]{Francesco Sanfilippo}
\author[c]{Alfredo Stanzione}

\affiliation[a]{Dipartimento di Fisica dell'Università di Pisa and INFN - Sezione di Pisa,\\
	Largo Pontecorvo 3, I-56127, Pisa, Italy}

\affiliation[b]{INFN - Sezione di Roma Tre,\\
	Via della Vasca Navale 84, I-00146, Rome, Italy}

\affiliation[c]{SISSA,\\
	Via Bonomea 265, I-34136, Trieste, Italy}

\emailAdd{massimo.delia@unipi.it}
\emailAdd{lorenzo.maio@phd.unipi.it}
\emailAdd{francesco.sanfilippo@infn.it}
\emailAdd{alfredo.stanzione@sissa.it}

\abstract{We discuss the phase diagram of QCD in the presence of a strong background magnetic field, providing numerical evidence, based on lattice simulations of QCD with $2+1$ flavours and physical quark masses, that the QCD crossover turns into a first order phase transition for large enough magnetic field, with a critical endpoint located between $eB=4$~GeV$^2$ (where we found an analytic crossover at a pseudo-critical temperature $T_c=(98\pm3)$~MeV) and $eB=9$~GeV$^2$ (where the measured critical temperature is $T_c=(63\pm5)$~MeV).}

\FullConference{
  41st International Conference on High Energy physics - ICHEP2022\\
  6-13 July, 2022\\
  Bologna, Italy
}

\begin{document}
\maketitle
\section{Introduction}\label{introduction}
Strongly interacting matter in a background magnetic field has been widely studied in the last decade~\cite{Andersen:2014xxa,Miransky:2015ava}. The interest is certainly due to the phenomenological relevance of this system. Indeed, strong magnetic fields interacting with hadronic matter can be found in three important contexts, whose theoretical models could be improved by these studies: magnetars~\cite{Duncan:1992hi}, heavy ion collisions~\cite{Skokov:2009qp} and the electroweak transition in the Early Universe~\cite{Grasso:2000wj}. Moreover, the intricate interplay between magnetic fields and non perturbative properties of QCD is extremely interesting from a theoretical point of view. For this reason, in this context, first principle lattice computations are extremely useful to get a full understanding on the system.

It is well known that a magnetic background affects the QCD vacuum, generating an anisotropy in the gluon field distribution, which alter chiral and confinement properties~\cite{Bonati:2014ksa,Bonati:2016kxj,Bonati:2018uwh,DElia:2021tfb}. In this work, we study the magnetic field effect at finite temperature. Previous studies highlighted that a background magnetic field affects the QCD phase transition, causing the crossover temperature, $T_c$, to decrease, and the crossover itself to strengthen: i.e. the jump in the observables becomes steeper and higher, when the magnetic field intensity, $B$, increases~\cite{Bali:2011qj,Endrodi:2015oba}. In particular, in~\cite{Endrodi:2015oba}, the author made a speculative proposal for the QCD phase diagram in a background magnetic field, predicting the appearance of a critical line with endpoint location at $eB_{CEP}\simeq10$~GeV$^2$ and $T_{CEP}\simeq105$~MeV. In this work we test such a prediction through simulations performed using tree-level improved Symanzik gauge action and the stout improved rooted staggered quark discretization, in unprecedented strong magnetic field for $2+1$ flavor QCD at the physical point, namely at $eB=4$ and $9$~GeV$^2$~\cite{DElia:2021yvk}.

This paper is organized as follows, in Section~\ref{the_transition_temperature} and~\ref{the_transition_nature}, we present our results on the (pseudo)critical temperature dependence on $B$ and the nature of the transition. In Section~\ref{conclusions} we draw our conclusions and propose an updated version of the QCD phase diagram in a background magnetic field.
\section{The transition temperature}\label{the_transition_temperature}
In the absence of external magnetic fields, at low temperatures quarks are confined into hadrons, chiral symmetry is spontaneously broken, i.e. the chiral condensate, $\displaystyle{\langle\overline{\psi}\psi\rangle}$, acquires a non zero value even in the limit of massless quarks. Raising the temperature above $T_c$, QCD matter undergoes a phase transition to a deconfined, chiral restored phase, where $\displaystyle{\langle\overline{\psi}\psi\rangle}$ vanishes. Thus, chiral condensate is the order parameter for such a transition. In the massive quark case, the phase transition switches to a crossover, and chiral condensate does not vanish in the hot phase. Nevertheless, it undergoes a smooth drop, hence it can be seen as a quasi order parameter, and it can be used to distinguish the different phases and to locate the transition temperature.

Chiral condensate brings cut-off dependent renormalization terms, which can be canceled subtracting its vacuum expectation value~\cite{Endrodi:2011gv}
\begin{equation}\label{eq:subtracted_cc}
	\Sigma^r_l(B,T) = \sum_{f=u,d}\langle\overline{\psi}\psi\rangle_f(B,T) - \langle\overline{\psi}\psi\rangle_f(B=0,T=0).
\end{equation}
Notice that \eqref{eq:subtracted_cc} differs from definition in~\cite{Endrodi:2011gv} by multiplicative factors, since we are only interested in the ratio $\displaystyle\Sigma^r_l(B,T)/\Sigma^r_l(0,0)$, where such overall factors cancel, as we computed $\Sigma^r_l(0,0)$ using the same cut-off we used for $\Sigma^r_l(B,T)$.

\begin{figure}
	\centering
	\includegraphics[width=0.4\textwidth]{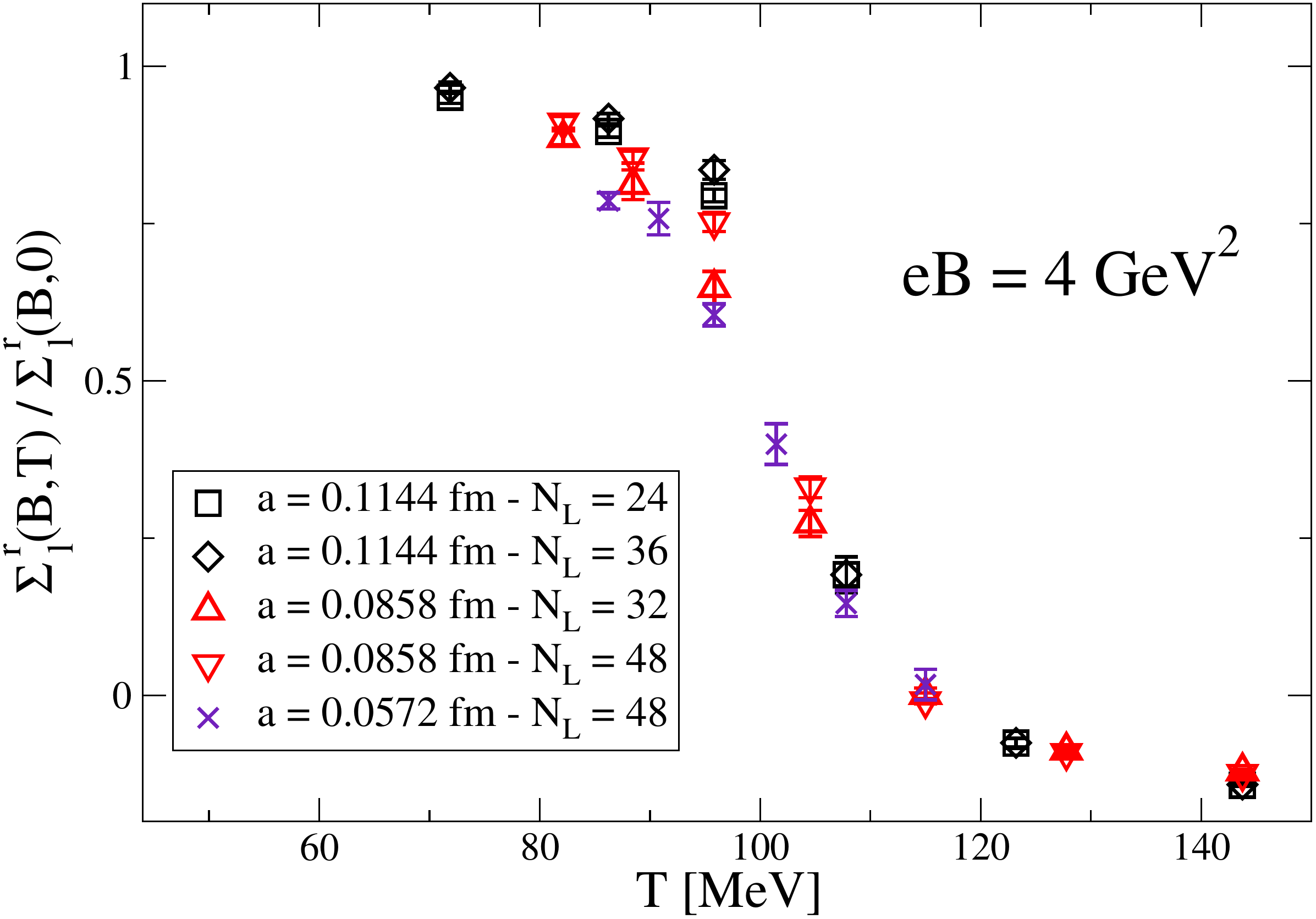}\qquad
	\includegraphics[width=0.4\textwidth]{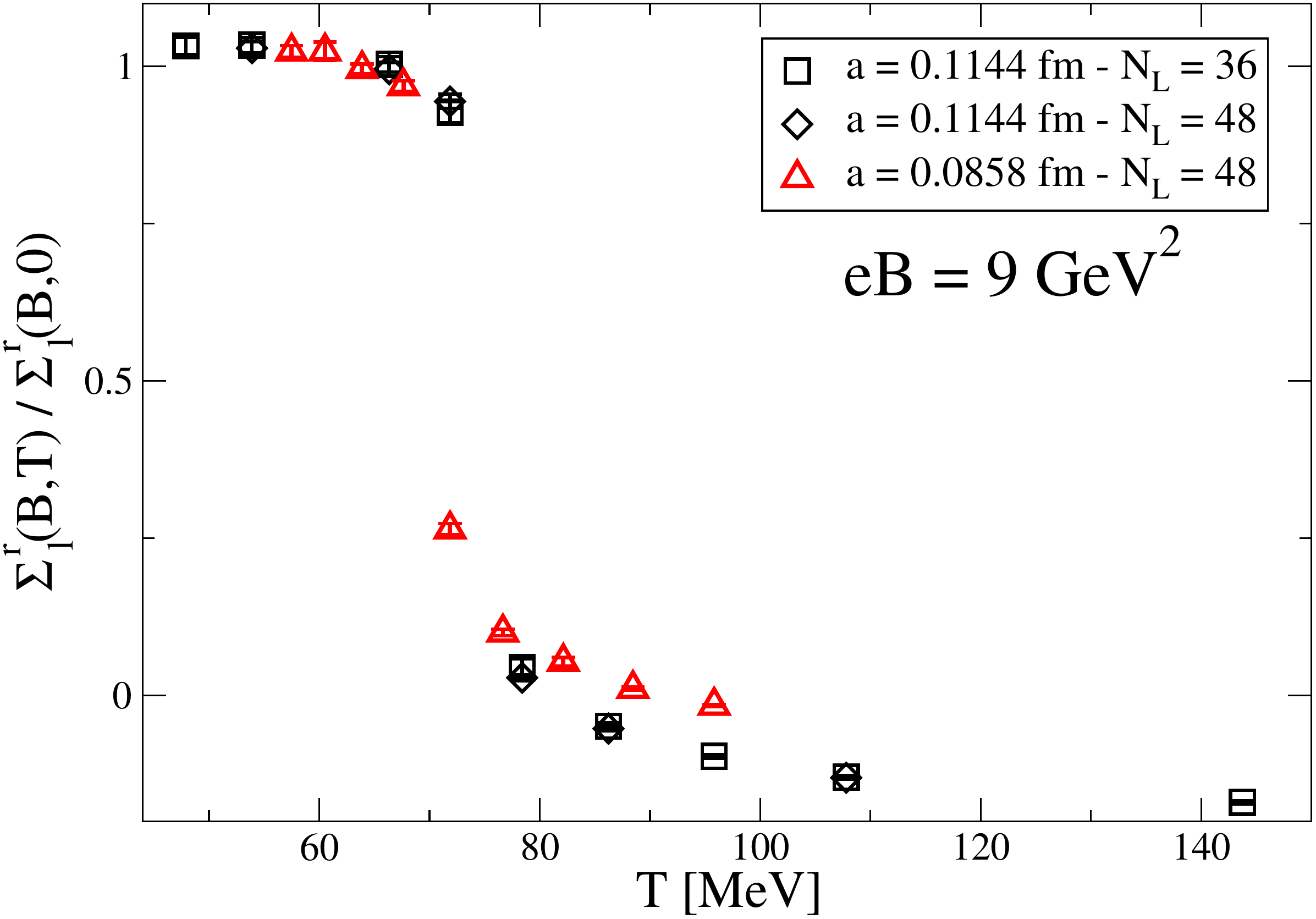}
	\caption{Renormalized chiral condensates, normalized with respect to their vacuum value, for $eB=4$ (left) and $9$~GeV$^2$ (right). In the latter, it can be observed the (pseudo)critical temperature drop, as well as the appearance of a gap in the chiral condensate.}
	\label{fig:chircond_vs_temp}
\end{figure}

In the two panels of Figure~\ref{fig:chircond_vs_temp} it is shown $\displaystyle\Sigma^r_l(B,T)/\Sigma^r_l(0,0)$ as a function of $T$, for the two studied values of $B$. It can be observed that the transition temperature decreases from $T_c\simeq100$~MeV to $T_c\simeq70$~MeV, when the magnetic field goes from $eB=4$~GeV$^2$ to $eB=9$~GeV$^2$. Moreover, a gap appears in the chiral condensate value across the transition temperature in the strongest magnetic field, as one could expect in the presence of a first order phase transition. 

\begin{figure}
	\centering
	\includegraphics[width=0.4\textwidth]{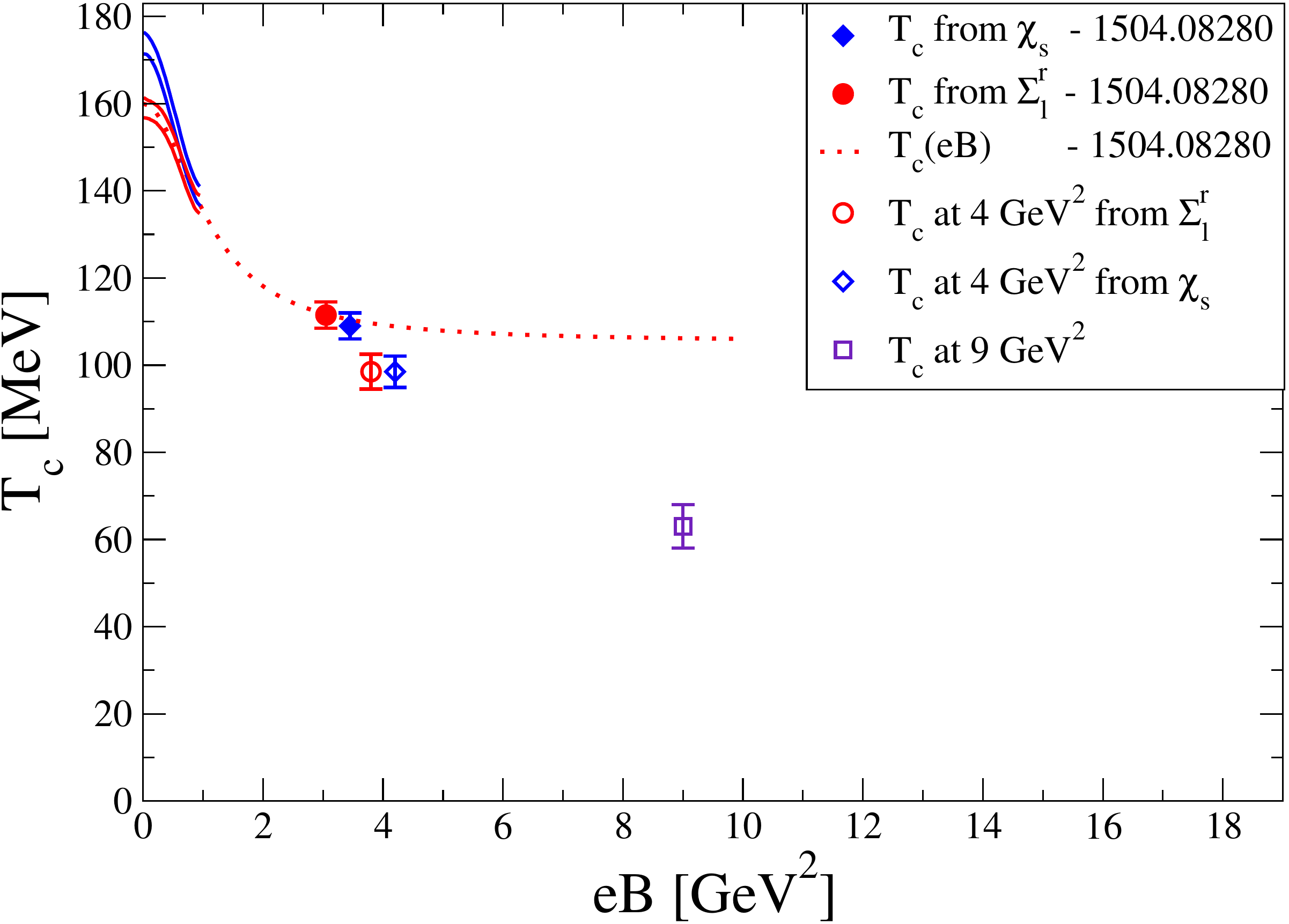}
	\caption{The measured transition temperatures as a function of the external magnetic field intensity in a first sketch of the updated phase diagram. Our results (empty markers) are compared to the predictions (dotted line) reported in~\cite{Endrodi:2015oba}, showing an unpredicted steady drop of $T_c$ as a function of the magnetic field.}
	\label{fig:critemp_vs_eb}
\end{figure}

In Figure~\ref{fig:critemp_vs_eb} we compare our results for the (pseudo)critical temperature $T_c(B)$, as a function of $B$, with the measurements and the predictions presented in~\cite{Endrodi:2015oba}. A fairly good agreement can be observed up to $eB=4$~GeV$^2$, region in which direct measurements were available; while the result we obtained for $eB=9$~GeV$^2$ is in disagreement with respect to the speculative prediction represented by the red dotted line. The steady drop we found could suggest that the critical temperature saturates to lower values as a function of $B$, but a steady drop to $0$ is not excluded.
\section{The transition nature}\label{the_transition_nature}
The smooth decrease of the chiral condensate in the left panel of Figure~\ref{fig:chircond_vs_temp}, is perfectly compatible with a crossover transition, while, as aforementioned, the gap appearing in the chiral condensate across the transition in the right panel of Figure~\ref{fig:chircond_vs_temp} is a smoking gun for a real phase transition. However, to unambiguously infer on the nature of the transition, it is mandatory to perform a finite size scaling~(FSS) analysis. Indeed, in a first order phase transition, in the thermodynamic limit, the chiral susceptibility, $\chi$, diverges at the critical temperature; in actual lattice simulations, such a divergence can only be observed in the scaling of $\chi(T_c)$ as a function of the spatial lattice size $\displaystyle L_S$, according to $\chi(L_S,T) = L_s^{\gamma/\nu} \phi((T - T_c) L_s^{1/\nu})$,
where $\nu = 1/3$ and $\gamma = 1$ are the first order critical indices. Thus, we studied the scaling behavior of the bare, disconnected up quark chiral susceptibility
\begin{equation}
	\chi_{u}^{disc} = \frac{1}{L_SN_t}\big[\langle{\overline{\psi}\psi}^2\rangle-{\langle\overline{\psi}\psi\rangle}^2\big],
\end{equation}
where $N_t$ is the lattice extension in euclidean time direction. To consider the full renormalized chiral susceptibility is irrelevant to our purpose, since $\chi_{u}^{disc}$ is expected to diverge itself at a real transition.
\begin{figure}
	\centering
	\includegraphics[width=0.393\textwidth]{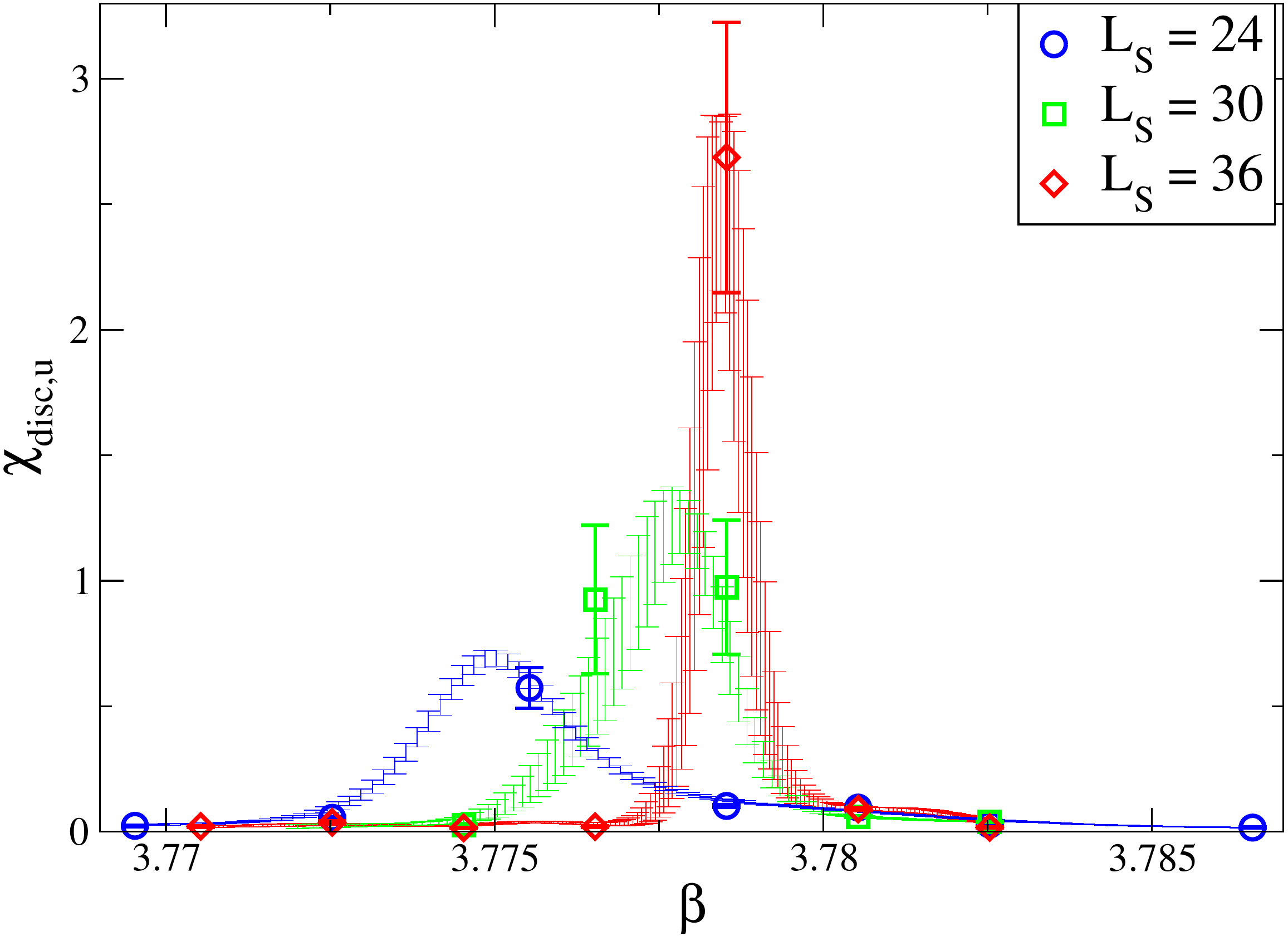}\qquad
	\includegraphics[width=0.407\textwidth, trim=0.1cm 0.12cm 0.4cm 0, clip]{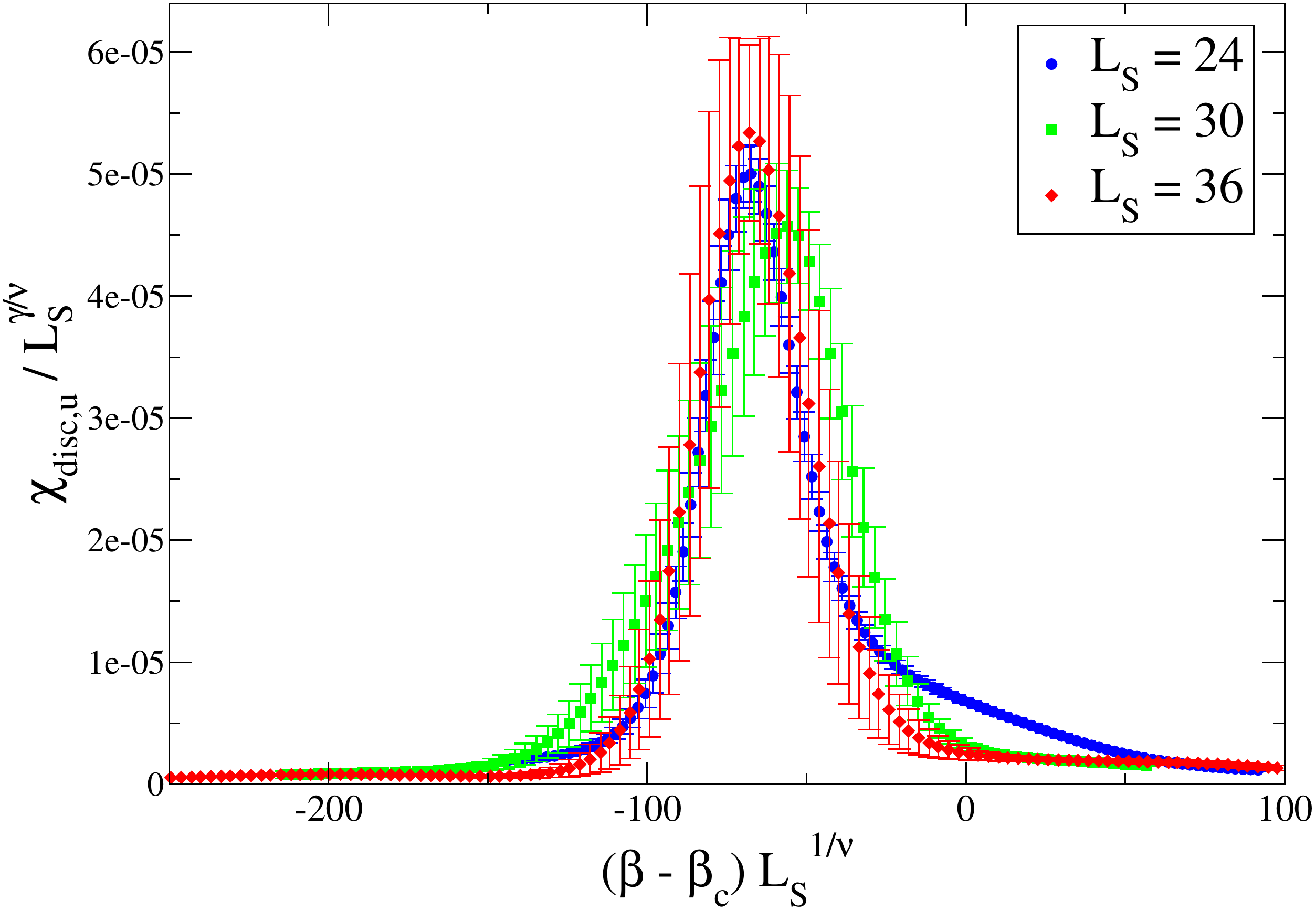}
	\caption{FSS analysis of the chiral susceptibility across the transition in the proximity of the constant physics line at $eB=9$~GeV$^2$ for the coarsest lattice spacing, where temperature is changed tuning $\beta$. The volume dependence indicates the presence of a real phase transition; 
	on the right hand side plot, we show the measurements rescaled using the first order critical exponents: the curves collapsing onto each other represent a strong evidence for a first order phase transition.}
	\label{fig:multihistogram}
\end{figure}

In Figure~\ref{fig:multihistogram}, the chiral susceptibility value across the phase transition is showed for three different lattice sizes $L_S$. In the left panel, the presence of a size scaling is obvious. In the right panel, data are rescaled using the first order critical exponents: the curves collapsing onto each other represent the strongest evidence for the presence of a first order phase transition.

\begin{figure}
	\centering
	\includegraphics[width=0.4\textwidth]{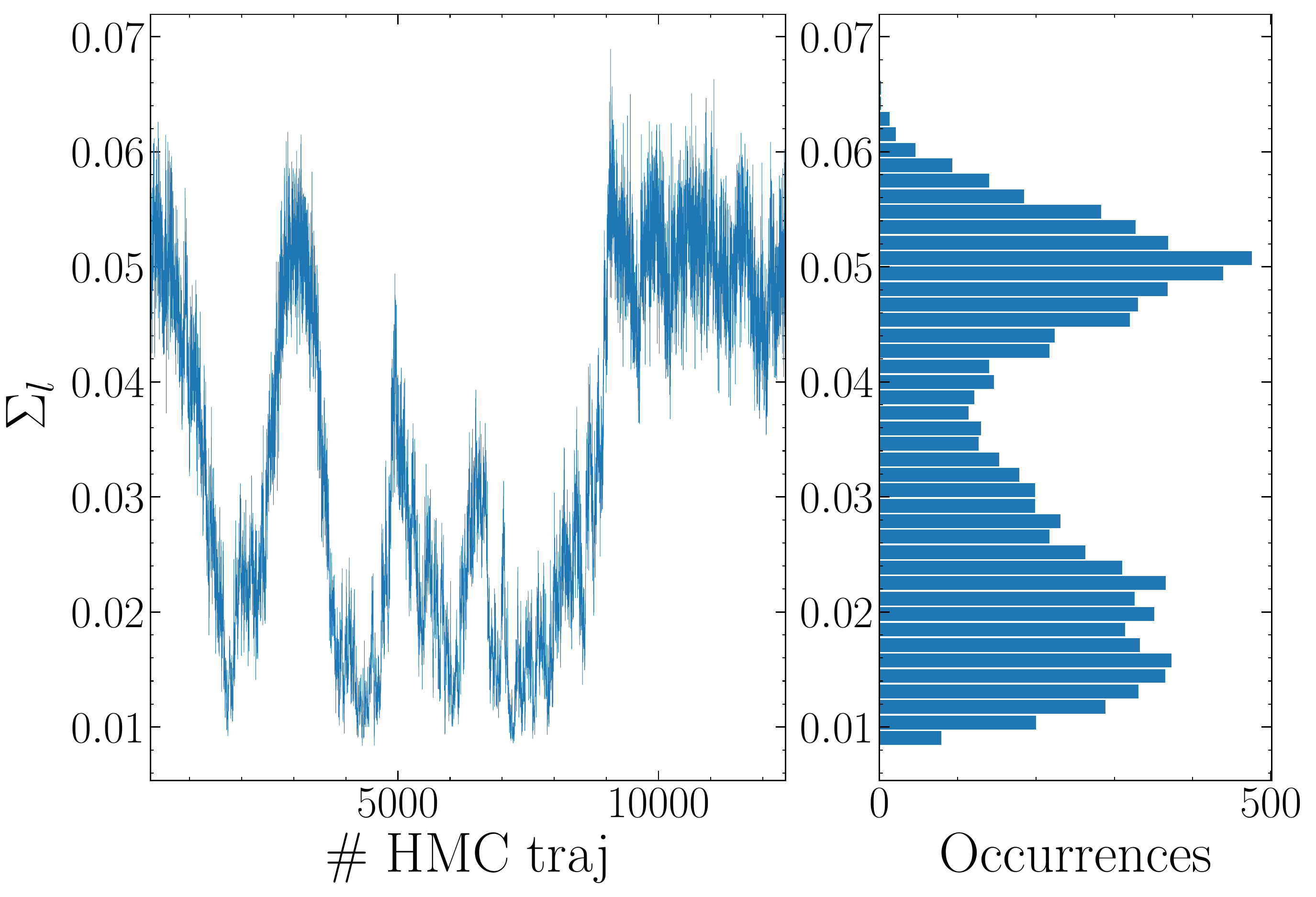}\qquad
	\includegraphics[width=0.4\textwidth]{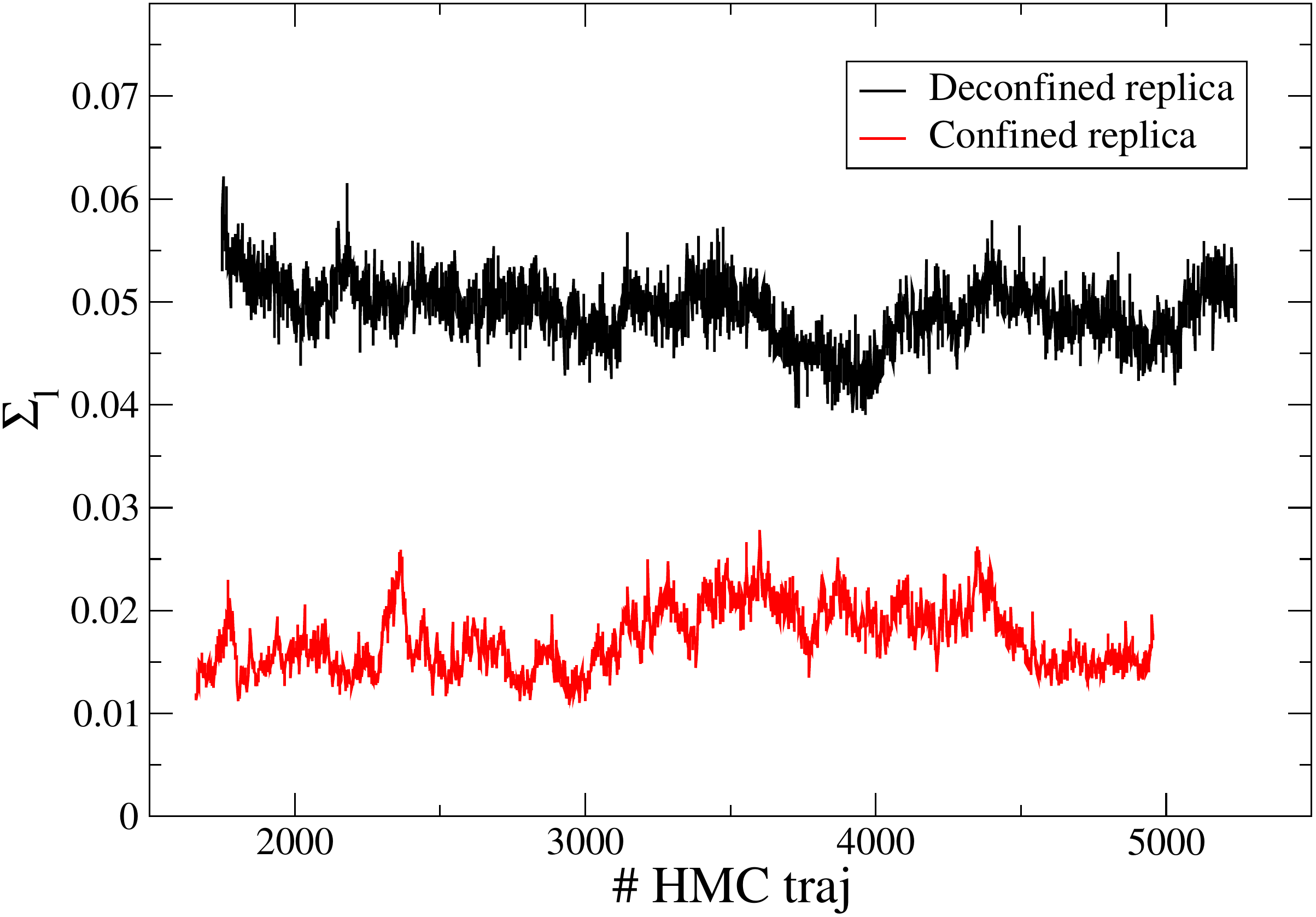}
	\caption{MCMC history of the chiral condensate computed on two samples next to the respective critical values of $\beta$ in different volumes. On the left hand side it is shown history of a $24^3$ lattice, oscillating between two different equilibrium states, clearly distinguishable in the chiral condensate probability distribution showed in the histogram. The right panel shows two simulations with an identical set of parameters, and starting points into two different phases, in a $36^3$ volume. The lowest tunnel probability with respect the left panel simulation is due to a larger physical volume, which enhances the potential barrier between the two different phases.}
	\label{fig:chiral_mc_histories}
\end{figure}

Besides FSS, Markov Chain Monte Carlo (MCMC) simulations, performed next to a first order phase transition, exhibit typical characterizing behavior. Because of the proximity, in the configuration space, of two different equilibrium points, the autocorrelation times of the simulations are expected to increase. This is due to the low effectiveness of the algorithm to tunnel between the two equilibrium states. The barrier separating the two phases is expected to grow as a function of the volume, diverging into the thermodynamic limit. In Figure~\ref{fig:chiral_mc_histories}, such a behavior can be observed: in the left panel, a simulation, performed at the critical values of the bare parameters, oscillates between two different equilibrium states, characterized by different values of the chiral condensate. In the right panel, it is presented the analogous situation in a bigger volume: the two lines represent two Monte Carlo simulations performed using the same set of parameters, but the respective starting points belong to two different phases. As expected, in the latter case, the tunneling probability is lower.

\section{Conclusions}\label{conclusions}
\begin{figure}
	\centering
	\includegraphics[width=0.4\textwidth]{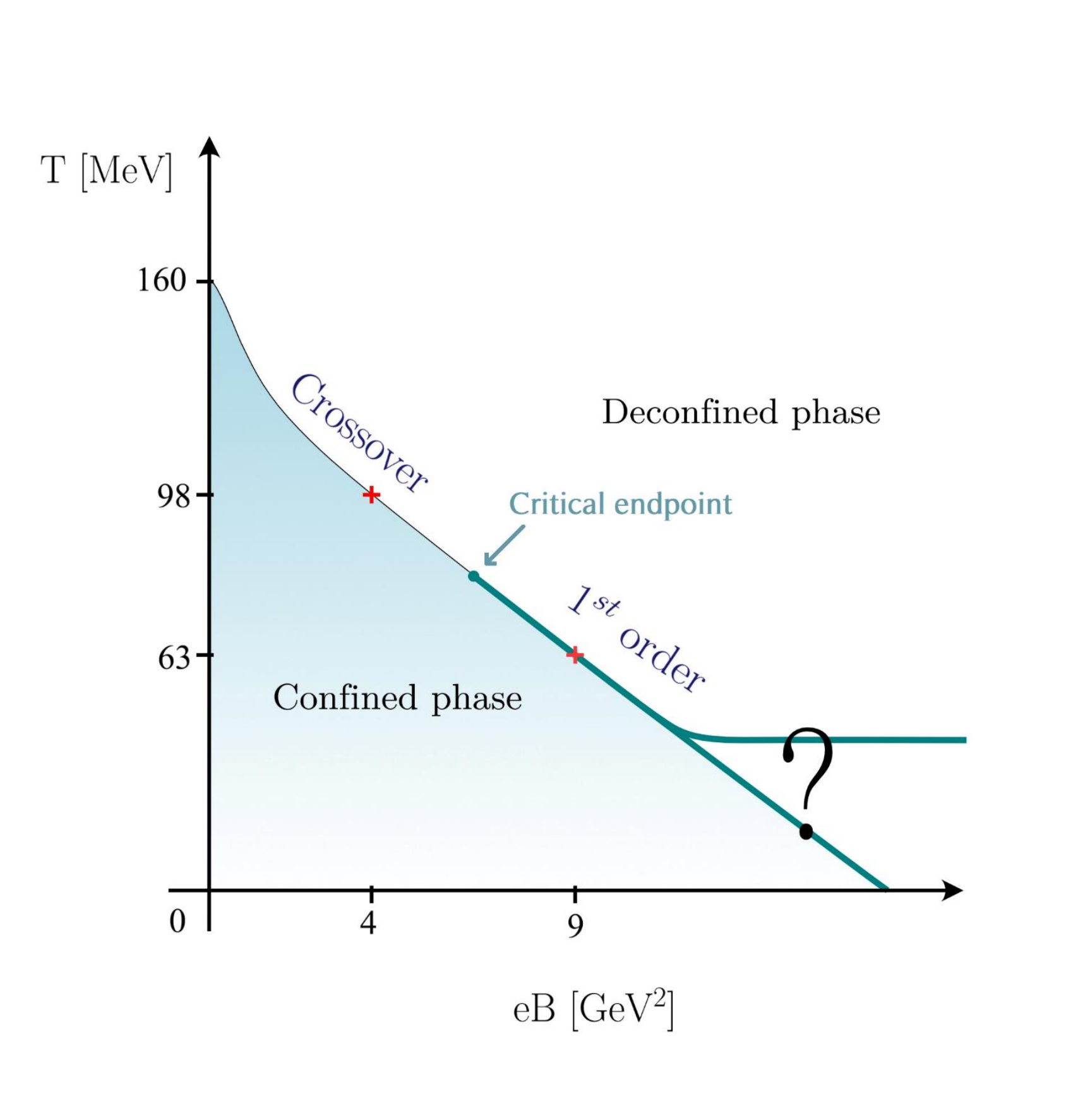}
	\caption{Updated QCD phase diagram in a background magnetic field, based on new evidences and speculations arising in this work. The (pseudo)critical temperature $T_c(eB)$ continues its drop as a function of $eB$. The critical end point is located in the temperature range $65$~MeV$<T_E<98$~MeV, which corresponds to the magnetic field range $4$~GeV$^2<eB_E<9$~GeV$^2$.}
	\label{fig:updated_pd}
\end{figure}

To summarize our results, we found a crossover transition in a $eB=4$~GeV$^2$ background magnetic field, at a temperature whose continuum extrapolation is $98$~MeV, and a first order phase transition in the presence of a $eB=9$~GeV$^2$ magnetic field, at a temperature that we estimate to range around $63$~MeV in the continuum limit. Thus, we infer the critical end point to be located on the straight line connecting these two points. Concerning the asymptotic behavior of $T_c(B)$, our measurements do not allow for any convincing prediction.
In Figure~\ref{fig:updated_pd}, relying on our findings, we propose an updated version of the $N_f=2+1$ QCD phase diagram, in the presence of a background magnetic field.

The present work is just a first exploration of the QCD properties in such strong magnetic fields, and demand for refinements. The first, obvious, follow-up consists in a better determination of the critical end point location, but such a task is numerically challenging, due to the aforesaid autocorrelation problem in the vicinity of a genuine phase transition, but also due to the low temperature and strong magnetic fields which require for big lattices to be properly simulated. However, there are at least two different ways to infer on its position: one could work on the high magnetic field side of the critical point, extrapolating the values of the magnetic field and temperature at which the two distinct equilibrium states collapse onto each other; alternatively, in the low magnetic field region, one could measure the thermodynamic variables, trying to detect the critical scaling behavior.

Finally, it would be interesting, also from a phenomenological point of view, to better characterize the two phases. In~\cite{DElia:2021yvk}, we studied the confining properties, finding that the chiral restored phase coincides, as expected, with a deconfined phase. We are currently studying electric conductivity in the deconfined phase, in the presence of such strong magnetic fields~\cite{Astrakhantsev:2019zkr}. Furthermore, it would be interesting to look at a more general view of the phase diagram, and explore the critical line behavior including a finite baryon density or a finite angular momentum.

\end{document}